\documentclass[aip,jcp,amsmath,amssymb,reprint,]{revtex4-2}

\usepackage[a-1b]{pdfx}
\usepackage{dcolumn} 
\usepackage{bm} 
\usepackage{graphicx}
\usepackage{multirow} 
\usepackage{pifont} 
\usepackage{epsfig}
\usepackage{amsmath} 
\usepackage{subfigure}
\usepackage{float}
\usepackage{xcolor}
\usepackage{color}
\usepackage{wrapfig}
\usepackage{braket}
\usepackage{threeparttable}
\usepackage{booktabs}
\usepackage{bm}
\usepackage{comment}
\usepackage{mathrsfs}
\usepackage{bbold}
\begin{document} 
\title{Enhanced diastereocontrol via strong light-matter interactions in an optical cavity}

\author{Nam Vu}
\affiliation{
             Department of Chemistry and Biochemistry,
             Florida State University,
             Tallahassee, FL 32306-4390}
\author{Grace M. McLeod}
\affiliation{
             Department of Chemistry and Biochemistry,
             Florida State University,
             Tallahassee, FL 32306-4390}  
             
\author{Kenneth Hanson}
\affiliation{
             Department of Chemistry and Biochemistry,
             Florida State University,
             Tallahassee, FL 32306-4390}     

\author{A. Eugene DePrince III}
\email{adeprince@fsu.edu}
\affiliation{
             Department of Chemistry and Biochemistry,
             Florida State University,
             Tallahassee, FL 32306-4390}


\begin{abstract}
The enantiopurification of racemic mixtures of chiral molecules is important for a range of applications. Recent work has shown that chiral group-directed photoisomerization is a promising approach to enantioenrich racemic mixtures of BINOL, but increased control of the diasteriomeric excess (\textit{de}) is necessary for its broad utility. Here we develop a cavity quantum electrodynamics (QED) generalization of time-dependent density functional theory and demonstrate computationally that strong light-matter coupling can alter the \textit{de} of chiral group-directed photoisomerization of BINOL. The relative orientation of the cavity mode polarization and the molecules in the cavity dictates the nature of the cavity interactions, which either enhance the \textit{de} of the (\textit{R})-BINOL diasteriomer (from 17\% to $\approx$ 40\%) or invert the favorability to the
(\textit{S})-BINOL derivative (to $\approx$ 34\% \textit{de}). The latter outcome is particularly remarkable because it indicates that the preference in diasteriomer can be influenced via orientational control, without changing the chirality of the directing group. We demonstrate that the observed effect stems from cavity-induced changes to the Kohn-Sham orbitals of the ground state.
\end{abstract}

\maketitle

\section{Introduction}

\begin{figure*}[!htpb]
	\centering
	\includegraphics[scale=0.3]{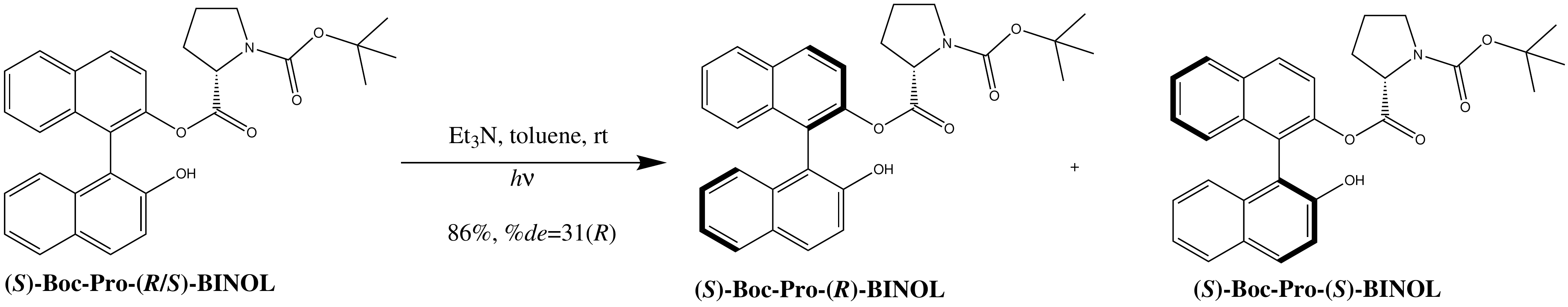}
	\caption{Enantioenrichment of (\textit{S})-Boc-Proline functionalized BINOL [(\textit{S})-Boc-Pro-(\textit{R}/\textit{S})-BINOL] by ESPT. The yield (86\%) and \textit{de} (31\%) correspond to those reported in Ref.~\citenum{Hanson19_1263}.}
	\label{FIG:BINOL_PURIFICATION}
\end{figure*}

Chiral molecules are ubiquitous in food additives, pharmaceuticals, catalysts, and elsewhere; the generation of enantiopure  molecules is thus critical for these applications.\cite{Trost04_5347} Molecules containing axial chirality like BINOL ([1,1'-binaphthalene]-2,2'-diol) and its derivatives are of particular interest because they are popular chiral ligands for a wide range of asymmetric catalytic reactions.\cite{Brunel05_857,Chen03_3155} Enantiopure BINOL ({\em i.e.}, either pure \textit{R} or \textit{S}) is typically obtained via chiral chromatography, strategic recrystallization, or direct asymmetric synthesis. However, separation methods often require large quantities of solvent or result in substantial loss of starting material ({\em i.e.} the undesired isomer), while synthetic means rely upon already enantiopure catalysts.\cite{Brunel05_857} 
Recently, chiral-group-directed photoisomerization was introduced as an alternative means of enantioenriching racemic mixtures of BINOL, and this strategy could theoretically result in 100\% yield and 100\% diastereomeric excess (\textit{de}).\cite{Hanson19_1263} Upon excitation in the presence of a base, BINOL is known to isomerize via an excited-state proton transfer (ESPT) mechanism.\cite{Solntsev09_227,Flegel08_161,Hanson19_580}
When one of its two -OH groups is functionalized with a chiral directing group [such as (\textit{S})-Boc-Proline, see Fig.~\ref{FIG:BINOL_PURIFICATION}] the isomerization is biased such that the \textit{de} at the photostationary state is dictated by the nature of this group and its impact on the energetics of the excited state diastereomers. While this approach shows promise, the best \textit{de} observed in Ref.~\citenum{Hanson19_1263} (63\%) was below the enantiopurity necessary for most applications ($>$95\%). Ultimately one would like to not only enhance this \textit{de} but also to exert some control over the chirality of the resulting product. Toward these aims, the present study explores how strong light-matter coupling 
can modulate the obtainable \textit{de} and diastereomeric preferences in ESTP-driven purification of BINOL derivatives.

Recently, there has been an explosion in interest harnessing strong light-matter interactions in optical cavities for chemical applications,\cite{Ebbesen16_2403,Narang18_1479,YuenZhou18_6325,Nori19_19} with a number of experimental and computational studies demonstrating various aspects of control over chemical transformations.\cite{George19_10635,Feist19_021057,Ebbesen19_615,Ebbesen16_11462,Ebbesen12_1592,Borjesson18_2273,Shegai18_eaas9552,KenaCohen19_eaax4482,Corni18_4688}  Cavity-induced changes to electronic structure could be particularly impactful in the areas of asymmetric synthesis and purification where even small changes in energy can have a large effect on the resulting enantiomeric/diastereomeric excess. Several recent computational studies have demonstrated that $>1$ kcal mol$^{-1}$ changes to spin-state splittings\cite{DePrince22_arxiv} or reaction barrier heights\cite{Flick22_4995, Rubio22_arxiv} can be realized via strong coupling of molecules to an optical cavity.
In the context of the ESPT-driven enantiopurification depicted in  Fig.~\ref{FIG:BINOL_PURIFICATION}, energy changes of this magnitude would result in dramatic changes to the observed \textit{de}. As an example, assuming that the \textit{de} reported in Ref.~\citenum{Hanson19_1263} are determined solely by the relative energies of the first excited states of the (\textit{S})-Boc-Pro-(\textit{R})-BINOL and (\textit{S})-Boc-Pro-(\textit{S})-BINOL diastereomers, the 63\% \textit{de} observed in that work would correspond to a roughly 0.9 kcal mol$^{-1}$ difference in energies in these states (see Eqs.~\ref{EQN:DE} and \ref{EQN:DELTA_E} below). A  $>95\%$ \textit{de} would require increasing this energy difference by roughly 1.3 kcal mol$^{-1}$. Given the magnitudes of energy changes predicted in other computational studies of cavity-bound molecules, it is reasonable to expect that sufficiently strong light-matter interactions could alter the relative energies of these states such that a $>95\%$ \textit{de} would be attainable via the ESPT mechanism considered here.

In this work, we use {\em ab initio} cavity quantum electrodynamics (QED) methods to explore how cavity interactions can influence the outcome of the ESPT-driven diastereomeric enrichment protocol shown in Fig.~\ref{FIG:BINOL_PURIFICATION}. We develop a cavity QED generalization of time-dependent density functional theory (TDDFT) for this problem in Sec.~\ref{SEC:THEORY} and outline the  details of our calculations in Sec.~\ref{SEC:COMPUTATIONAL_DETAILS}. In Sec.~\ref{SEC:RESULTS}, we apply QED-TDDFT to this diastereomeric enrichment problem, and we find that strong light-matter coupling can drive the \textit{de} toward {\em either} diastereomer, depending on orientation of the molecule relative to the cavity mode polarization. After some concluding remarks in Sec.~\ref{SEC:CONCLUSIONS}, a complete derivation of the QED-TDDFT approach that we employ can be found in Appendices A and B.

\section{Theory}
\label{SEC:THEORY}

Computational cavity QED studies often use simple model Hamiltonians\cite{Cummings63_89,Cummings68_379} that describe interactions between quantized radiation modes and few-level quantum emitters.  A more rigorous description of molecular degrees of freedom can be obtained from {\em ab initio} cavity QED approaches, which resemble familiar electronic structure methods, but are generalized to describe both electron-electron and electron-photon interactions. Examples of calculations performed using cavity QED extensions of density functional theory,\cite{Bauer11_042107,Rubio14_012508,Tokatly13_233001,Rubio17_3026,Rubio17_113036,Tokatly18_235123,Rubio15_093001,Rubio18_992,Rubio19_4883,Appel19_225,Narang20_094116,Rubio20_508,Narang21_104109,Shao21_064107,Shao22_124104,Rubio21_41} coupled-cluster theory,\cite{Koch20_041043,Manby20_023262,DePrince21_094112,DePrince22_2111,Flick22_4995,Koch21_094113,Flick21_9100,Koch22_2203,Rubio22_JCP_0}  configuration interaction,\cite{Foley22_154103}  or reduced-density-matrix methods\cite{DePrince22_arxiv} are becoming increasingly commonplace. 
In this work, we adopt a QED-TDDFT formalism that most closely resembles the Gaussian-basis formalism described in Ref.~\citenum{Shao21_064107}. 
A detailed derivation of working equations for QED-TDDFT can be found in that work, and we present our own derivation, which results in slightly different equations, in Appendix B. In this section, our aim is to describe the approach with enough detail such that slight differences between the formalism outlined in Ref.~\citenum{Shao21_064107} and that which we use can be understood.

Interactions between electronic degrees of freedom and quantized radiation fields associated with an optical cavity can be described by the Pauli-Fierz (PF) Hamiltonian.\cite{Spohn04_book,Rubio18_0118} We limit our considerations to a cavity that supports a single  photon mode, and we express this Hamiltonian in the length gauge and under the dipole and cavity Born-Oppenheimer approximations as
\begin{eqnarray}
    \label{EQN:HPF}
    \hat{H}_{\rm PF}&=&\hat{H}_{\rm e}+\omega_{\rm cav}\hat{b}^{\dagger}\hat{b} -\sqrt{\frac{\omega_{\rm cav}}{2}}(\bm{\lambda} \cdot \bm{\hat{\mu}})(\hat{b}^{\dagger}+\hat{b}) \nonumber \\ 
    &+&\frac{1}{2}(\bm{\lambda} \cdot \bm{\hat{\mu}})^2
\end{eqnarray}   
Here, the first two terms are the usual electronic Hamiltonian ($\hat{H}_{\rm e}$) and the Hamiltonian for the photon mode; $\omega_{\rm cav}$ is the fundamental frequency associated with this mode, and $\hat{b}^\dagger$ and $\hat{b}$ represent bosonic creation and annihilation operators, respectively. The third and fourth terms in Eq.~\ref{EQN:HPF} represent the bilinear coupling between the electron and photon degrees of freedom and the dipole self-energy, respectively. The symbol ${\bm{\hat{\mu}}}$ represents the total molecular dipole operator (electronic plus nuclear, {\em i.e.}, ${\bm{\hat{\mu}}} = {\bm{\hat{\mu}}}_{\rm e} + {\bm{\hat{\mu}}}_{\rm n}$), and the coupling vector, ${\bm \lambda}$, parametrizes the strength of the photon-electron interactions. 
We are interested in single-molecule coupling, in which case we take ${\bm \lambda} = \lambda {\bm u}$, where ${\bm u}$ is a unit vector describing the polarization of the cavity mode, and the magnitude of the coupling vector, $\lambda$, relates to the effective cavity mode volume as \cite{Feist19_021057}
\begin{equation}
    \label{eq:lambda}
    \lambda=\sqrt{\frac{1}{\epsilon_0 V_{\rm eff}}}
\end{equation}
Here, $\epsilon_0$ is the permittivity of free-space. At this point, we can note one difference between the present formalism and that outlined in Ref.~\citenum{Shao21_064107}. In Ref.~\citenum{Shao21_064107}, the expectation value of the dipole operator enters Eq.~\ref{EQN:HPF}, rather than the dipole operator itself; in that case, as described below, cavity interactions do not perturb the ground-state Kohn-Sham orbitals. On the other hand, with the Hamiltonian in Eq.~\ref{EQN:HPF}, the Kohn-Sham orbitals can relax to account for the presence of the cavity. For this reason, we refer to QED-TDDFT based on the formalisms outlined in Ref.~\citenum{Shao21_064107} and herein as ``unrelaxed'' and ``relaxed'' QED-TDDFT, respectively.

Similar to the case in Kohn-Sham DFT, the ground-state in QED-DFT maps onto a non-interacting reference function of the form
\begin{equation}
    |\Psi\rangle = |0^{\rm e}\rangle \otimes |0^{\rm p}\rangle
\end{equation}
where $|0^{\rm e}\rangle$ refers to a Kohn-Sham determinant of electronic spin orbitals, and $|0^{\rm p}\rangle$ represents a zero-photon state. These functions can be determined via a modified Roothaan-Hall procedure: (i) $|0^{\rm e}\rangle$ can be determined as the Kohn-Sham determinant that minimizes the electronic energy, given a fixed $|0^{\rm p}\rangle$, and (ii) $|0^{\rm p}\rangle$ can be determined as the lowest-energy eigenfunction of $\langle \hat{H}_{\rm PF}\rangle_{\rm e}$, where the subscript ``e'' indicates that we have integrated out the electronic degrees of freedom. For the first step, electron correlation and exchange effects can be accounted for using standard density functional approximations. For the second step, $\langle \hat{H}_{\rm PF}\rangle_{\rm e}$ can be expanded in a basis of photon-number states and diagonalized to find $|0_{\rm p}\rangle$. 
This procedure can be repeated until self-consistency. Herein lies the primary difference between the relaxed and unrelaxed QED-TDDFT formalisms: in the former, the ground state electronic orbitals are determined in the presence of the cavity interaction and dipole self-energy terms, whereas those in the latter are not. We note that neither approach considers any electron-photon correlation effects (such as those captured by electron-photon correlation functionals described in Refs.~\citenum{Rubio15_093001,Rubio18_992,Flick21_arxiv}).

An equivalent representation of relaxed ground-state QED-DFT involves representing the problem within the coherent-state basis,\cite{Koch20_041043} which is the basis that diagonalizes $\langle \hat{H}_{\rm PF}\rangle_{\rm e}$. In this way, we avoid the need to solve the second step of the modified Roothaan-Hall procedure described above. Rather, we solve only the electronic problem with a modified Hamiltonian of the form
\begin{eqnarray}
    \label{EQN:HPF_COHERENT}
    \hat{H}_{\rm PF}&=&\hat{H}_{\rm e}+\omega_{\rm cav}\hat{b}^{\dagger}\hat{b} -\sqrt{\frac{\omega_{\rm cav}}{2}}(\bm{\lambda} \cdot [\bm{\hat{\mu}} - \langle\bm{\hat{\mu}}\rangle] )(\hat{b}^{\dagger}+\hat{b}) \nonumber \\ 
    &+&\frac{1}{2}(\bm{\lambda} \cdot [\bm{\hat{\mu}} - \langle\bm{\hat{\mu}}\rangle])^2
\end{eqnarray}  
where $\langle\bm{\hat{\mu}}\rangle$ represents the expectation value of the molecular dipole with respect to the Kohn-Sham determinant. Additional details regarding ground-state QED-DFT can be found in Appendix A.

For excitation energies of cavity-bound molecules, we use a QED generalization of TDDFT; a derivation of this approach can be found in Appendix B. The resulting generalized eigenvalue problem is\cite{Shao21_064107}
\begin{widetext}
\begin{equation}
\label{EQN:QED_TDDFT_EQUATIONS}
\begin{pmatrix}
\bm{A+\Delta} &\bm{B+\Delta^\prime} & \bm{ g^\dagger} & \bm{ g^\dagger} \\
\bm{B+\Delta^\prime} &\bm{A+\Delta} & \bm{ g^\dagger} & \bm{ g^\dagger}\\
\bm{ g} & \bm{ g} &\omega_{\rm cav} &0  \\
\bm{ g} & \bm{ g} &0 &\omega_{\rm cav}\\
\end{pmatrix}
\begin{pmatrix}
X  \\
Y \\
M \\
N\\
\end{pmatrix}
=\Omega 
\begin{pmatrix}
\bm{1} &0 & 0 & 0 \\
0 &\bm{-1} & 0 & 0\\
0 & 0 &\bm{1} &0  \\
0 & 0 &0 &\bm{-1}  \\
\end{pmatrix}
\begin{pmatrix}
X  \\
Y \\
M \\
N\\
\end{pmatrix}
\end{equation}
\end{widetext}
On the left-hand side of Eq.~\ref{EQN:QED_TDDFT_EQUATIONS}, ${\bm A}$ and ${\bm B}$ are the same matrices that arise in the usual (non-QED) TDDFT problem,
${\bm \Delta}$ and ${\bm \Delta^\prime}$ represent dipole self-energy contributions, and ${\bm g}$ arises from the bilinear coupling term. Explicit expressions for these quantities (for the case of the random phase approximation) can be found in Appendix B. 
On the right-hand side of Eq.~\ref{EQN:QED_TDDFT_EQUATIONS}, the symbol $\Omega$ represents an excitation energy, and the vectors $X$, $Y$, $M$, and $N$ parametrize the corresponding QED-TDDFT excited state 
(see Eq.~\ref{EQN:RPA_EXPANSION} in Appendix B) and contain amplitudes corresponding to electronic excitations, electronic de-excitations, photon creation, and photon annihilation, respectively. Additional details can be found in Appendix B.

\section{Computational Details}
\label{SEC:COMPUTATIONAL_DETAILS}

The QED-TDDFT method was implemented in \texttt{hilbert},\cite{hilbert1} which is a plugin to the \textsc{Psi4}\cite{Sherrill20_184108} electronic structure package. 
QED-TDDFT calculations on deprotonated (\textit{S})-Boc-Pro-(\textit{R}/\textit{S})-BINOL (charge = -1) molecules were performed using the 6-31G(d,p) basis set, using density-fitted two-electron integrals and the cc-pVDZ-JKFIT auxiliary basis set. Geometries for deprotonated (\textit{S})-Boc-Pro-(\textit{R}/\textit{S})-BINOL were taken from Ref.~\citenum{Hanson19_1263}, which were optimized for the first singlet excited state in either molecule, at the B3LYP/6-31G(d,p) level of theory. As mentioned in Sec.~\ref{SEC:THEORY}, we use standard density functional approximations (SVWN3, PBE, and B3LYP) from electronic structure theory in the QED-TDDFT calculations, and we neglect electron-photon correlation effects.

In all QED-TDDFT calculations performed in this work, we consider a single-mode cavity, and details regarding the relative orientation of the cavity mode axis and the molecule can be found in Sec.~\ref{SEC:RESULTS}. We consider coupling strengths in the range $\lambda = 0.01$ to $\lambda = 0.05$, which correspond to effective mode volumes (Eq.~\ref{eq:lambda}) as large as $\approx$ 18.6 nm$^3$ (for $\lambda = 0.01$)  or as small as $\approx$ 0.74 nm$^3$ (for $\lambda = 0.05$). We consider multiple values for the fundamental frequency of the cavity mode ($\omega_{\rm cav}$); additional details can be found in Sec.~\ref{SEC:RESULTS}.

\section{Results and Discussion}
\label{SEC:RESULTS}

\begin{figure}[!htpb]
	\includegraphics[angle=270, scale=0.3]{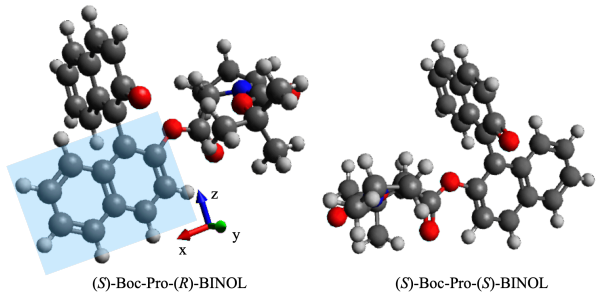}
	\caption{Orientation of deprotonated (\textit{S})-Boc-Pro-(\textit{R}/\textit{S})-BINOL used on the QED-TDDFT-B3LYP calculations.  }\label{FIG:MOLECULAR_ORIENTATION}
\end{figure}

As a representative example of cavity-enhanced ESPT-mediated enantioenrichment of BINOL, we consider the case of (\textit{S})-Boc-Proline functionalized BINOL, which has previously been examined at the TDDFT [B3LYP / 6-31G(d,p)] level of theory.\cite{Hanson19_1263} The present QED-TDDFT calculations take the molecules to be oriented as depicted in Fig.~\ref{FIG:MOLECULAR_ORIENTATION}, with the $xz$ plane defined by the plane of the (\textit{S})-Boc-Pro functionalized naphthol moiety. Given this configuration, we have considered cavity modes polarized along each cartesian axis ($x$, $y$, and $z$), as well as along the axis defined by the molecular dipole moment for each molecule. We consider three different fundamental frequencies. First,  we take $\omega_{\rm cav} = \omega_{R} = 1.37630$ eV, which is resonant with the energy of the first excited state of deprotonated (\textit{S})-Boc-Pro-(\textit{R})-BINOL, as predicted by B3LYP / 6-31G(d,p). Second, we use $\omega_{\rm cav} = \omega_{S} = 1.33662$ eV, which is resonant with the energy of the first excited state of deprotonated (\textit{S})-Boc-Pro-(\textit{S})-BINOL, as predicted by the same level of theory. Lastly, we use $\omega_{\rm cav} = 10.0$ eV, which serves as a non-resonant case.

Table \ref{TAB:B3LYP_RELAXED} shows relaxed QED-TDDFT predictions of the \textit{de} for ESPT-mediated enantioenrichment of (\textit{S})-Boc-Pro-(\textit{R}/\textit{S})-BINOL, using the B3LYP functional and the 6-31G(d,p) basis set. 
The \textit{de} was determined as outlined in Ref.~\citenum{Hanson19_1263}, as
\begin{equation}
\label{EQN:DE}
    {de} = \bigg | \frac{f_1 - f_2}{f_1 + f_2} \bigg | \times 100 \%
\end{equation}
where $f_1$ represents the fraction of the diastereomer that is in excess, $f_2$ = 1 - $f_1$ represents the fraction of the other diastereomer, and these fractions are related to the energy difference between the first excited electronic states of the deprotonated diastereomers by
\begin{equation}
\label{EQN:DELTA_E}
    \Delta E =E_2-E_1= -RT~{\rm ln} \bigg ( \frac{1-f_1}{f_1} \bigg )
\end{equation}
Here, the thermal, rotational, and vibrational contributions to the total internal energy and to the entropy were assumed to be equal for both diasteromers, and, thus, $\Delta G$ for the photoisomerization is simply the difference in the electronic energies of the first excited electronic states. The temperature, $T$, is taken to be 298.15 K. We note that this simple analysis reproduces the correct qualitative observation that (\textit{S})-Boc-Pro-(\textit{R})-BINOL is the preferred diasteromer, with TDDFT predicting a \textit{de} of 17\% and experiment giving 31\%.\cite{Hanson19_1263}  

The data in Table \ref{TAB:B3LYP_RELAXED} indicate that the computed diasteriomeric excesses are sensitive to cavity interactions, and changes to the \textit{de} depend strongly on the relative orientations of the molecule and the cavity mode axis. For example, with a $z$-polarized cavity mode, the \textit{de} increases by 23--24\%, from the cavity-free preference for the (\textit{S})-Boc-Pro-(\textit{R})-BINOL diasteromer of 17\% to as much as a 41\% excess of the same diastereomer (with a coupling strength of $\lambda = 0.05$ atomic units and $\omega_{\rm cav} = \omega_R$ or $\omega_S$). 
Interestingly, the preference for the (\textit{S})-Boc-Pro-(\textit{R})-BINOL can be reversed when the cavity mode is polarized along the molecular dipole moment, in which case we observe as much as 34\% \textit{de} of (\textit{S})-Boc-Pro-(\textit{S})-BINOL (again, with a coupling strength of $\lambda = 0.05$ atomic units). For a $x$- and $y$-polarized cavity modes, we find only modest decreases in the preference for the (\textit{S})-Boc-Pro-(\textit{R})-BINOL diastereomer, for all choices of $\omega_{\rm cav}$. 

	\begin{table}
		\begin{threeparttable}
			\caption {Computed diastereomeric excess for (\textit{S})-Boc-Pro-(\textit{R}/\textit{S})-BINOL from relaxed QED-TDDFT calculations. The \textit{de} was determined according to the relative energies of the first electronic excited states of deprotonated (\textit{S})-Boc-Pro-(\textit{R}/\textit{S})-BINOL. } \label{TAB:B3LYP_RELAXED}
			\footnotesize\centering
			\setlength{\tabcolsep}{0em}
			\begin{tabular*}{\linewidth}{@{\extracolsep{\fill}}
					lccccc}
			
			& & \multicolumn{3}{c}{mode polarization} \\
			\cline{3-5}
			&$\lambda$& $x$ & $y$ & $z$ &dipole\\
\midrule
                             &0.00   & 17(\textit{R}) & 17(\textit{R})& 17(\textit{R}) &17(\textit{R})  \\
\midrule
                             &0.01   & 17(\textit{R}) & 17(\textit{R}) & 18(\textit{R}) & 14(\textit{R})   \\
$\omega_{\rm cav} =\omega_R$ &0.02   & 16(\textit{R}) & 16(\textit{R}) & 21(\textit{R}) &  7(\textit{R})   \\
                             &0.03   & 15(\textit{R}) & 14(\textit{R}) & 27(\textit{R}) &  5(\textit{S})   \\
                             &0.04   & 13(\textit{R}) & 13(\textit{R}) & 34(\textit{R}) & 19(\textit{S})   \\
                             &0.05   & 11(\textit{R}) & 12(\textit{R}) & 41(\textit{R}) & 34(\textit{S})   \\

\midrule
                             &0.01   & 17(\textit{R}) &17(\textit{R})  & 18(\textit{R}) & 15(\textit{R})   \\
$\omega_{\rm cav} =\omega_S$ &0.02   & 16(\textit{R}) &16(\textit{R})  & 21(\textit{R}) &  6(\textit{R})   \\
                             &0.03   & 15(\textit{R}) & 14(\textit{R}) & 27(\textit{R}) &  5(\textit{S})   \\
                             &0.04   & 13(\textit{R}) & 13(\textit{R}) & 34(\textit{R}) & 19(\textit{S})   \\
                             &0.05   & 11(\textit{R}) & 12(\textit{R}) & 41(\textit{R}) & 34(\textit{S})   \\

\midrule
                             &0.01   & 17(\textit{R}) & 17(\textit{R}) & 18(\textit{R}) & 14(\textit{R})   \\
$\omega_{\rm cav} = 10.0$ eV &0.02   & 16(\textit{R}) & 16(\textit{R}) & 21(\textit{R}) &  6(\textit{R})   \\
                             &0.03   & 15(\textit{R}) & 14(\textit{R}) & 26(\textit{R}) &  5(\textit{S})   \\
                             &0.04   & 13(\textit{R}) & 13(\textit{R}) & 33(\textit{R}) & 19(\textit{S})   \\
                             &0.05   & 11(\textit{R}) & 12(\textit{R}) & 40(\textit{R}) & 34(\textit{S})  \\

		\end{tabular*}
	\end{threeparttable}
\end{table}

It is notable that the same conclusions can be drawn from the data in Table \ref{TAB:B3LYP_RELAXED} for each chosen $\omega_{\rm cav}$, as the observed \textit{de} for $\omega_{\rm cav} = \omega_R$, $\omega_S$, and $10.0$ eV are essentially identical. This insensitivity suggests that the  changes to the \textit{de} we observe derive from dipole-self-energy-induced modifications to either the ground or excited electronic states. 
For the ground state, when the Hamiltonian is represented in the coherent-state basis, the energy (at the mean-field level) is independent of the cavity mode frequency (see Eq.~\ref{EQN:QED_DFT_ENERGY} in Appendix A), and all cavity effects stem from dipole self-energy. As for the excited-states, 
from a model Hamiltonian perspective, we would expect interactions between the electronic excited states and the photon degrees of freedom to be negligible in the case that the transition dipole moments for the electronic states are small. Indeed, the data in Table \ref{TAB:TDPM} support this expectation; transition dipole moments for these states are small in all cartesian directions, and the oscillator strengths are only on the order of 10$^{-5}$--10$^{-4}$. These small transition moments are consistent with the argument that the dipole self-energy must be responsible for the cavity effects we observe.

	\begin{table}
		\begin{threeparttable}
			\caption {Transition dipole moments (units of $e a_0$) and oscillator strengths ($f$) corresponding to the first electric excited states of (\textit{S})-Boc-Pro-(\textit{R}/\textit{S})-BINOL calculated at the TDDFT [B3LYP/6-31G(d,p)] level of theory.  } \label{TAB:TDPM}
			\footnotesize\centering
			\setlength{\tabcolsep}{0em}
			\begin{tabular*}{\linewidth}{@{\extracolsep{\fill}}
					lcccc}
			&$\mu_x$& $\mu_y$ & $\mu_z$ & $f$  \\
			\midrule
			(\textit{S})-Boc-Pro-(\textit{R})-BINOL     & \phantom{-}0.0049  & -0.0149  & -0.0316   & 4.2 $\times 10^{-5}$ \\ 
			(\textit{S})-Boc-Pro-(\textit{S})-BINOL     &-0.0178  &  \phantom{-}0.0109  & -0.0600   & 1.3 $\times 10^{-4}$ \\ 

		\end{tabular*}
	\end{threeparttable}
\end{table}

We can confirm the importance of the dipole self-energy term's influence on the ground state by evaluating the \textit{de} with energies derived from unrelaxed QED-TDDFT calculations. Table \ref{TAB:B3LYP_UNRELAXED} provides these data for the off-resonance case of $\omega_{\rm cav}$ = 10.0 eV. As in Table \ref{TAB:B3LYP_RELAXED}, these data were generated using the B3LYP functional and the 6-31G(d,p) basis set, and we consider four different cavity mode polarization axes. We find that the \textit{de} is completely insensitive to cavity interactions, remaining at roughly 17\% for all coupling strengths and cavity mode polarization axes. These results stand in stark contrast to those of Table \ref{TAB:B3LYP_RELAXED} and highlight the importance of ground-state orbital relaxation effects in the strong coupling regime. 

	\begin{table}
		\begin{threeparttable}
			\caption {Computed diastereomeric excess for (\textit{S})-Boc-Pro-(\textit{R}/\textit{S})-BINOL from unrelaxed QED-TDDFT calculations. The \textit{de} was determined according to the relative energies of the first electronic excited states of deprotonated (\textit{S})-Boc-Pro-(\textit{R}/\textit{S})-BINOL. } \label{TAB:B3LYP_UNRELAXED}
			\footnotesize\centering
			\setlength{\tabcolsep}{0em}
			\begin{tabular*}{\linewidth}{@{\extracolsep{\fill}}
					lccccc}
			
			& & \multicolumn{3}{c}{mode polarization} \\
			\cline{3-5}
			&$\lambda$& $x$ & $y$ & $z$ &dipole\\
\midrule
                              &0.00   & 17(\textit{R})  & 17(\textit{R}) & 17(\textit{R}) & 17(\textit{R})   \\
\midrule                        
                              &0.01   & 17(\textit{R})  & 17(\textit{R}) & 17(\textit{R}) & 17(\textit{R})  \\
$\omega_{\rm cav} = 10.0$ eV  &0.02   & 17(\textit{R})  & 17(\textit{R}) & 17(\textit{R}) & 17(\textit{R})  \\
                              &0.03   & 17(\textit{R})  & 17(\textit{R}) & 17(\textit{R}) & 17(\textit{R}) \\
                              &0.04   & 17(\textit{R})  & 17(\textit{R}) & 17(\textit{R}) & 17(\textit{R}) \\
                              &0.05   & 17(\textit{R})  & 17(\textit{R}) & 17(\textit{R}) & 17(\textit{R}) \\
		\end{tabular*}
	\end{threeparttable}
\end{table}

Lastly, we explore whether the choice of exchange-correlation (XC) functional has any effect on cavity-induced changes to the \textit{de}. The formulation of QED-TDDFT that we use lacks any photon-electron XC functional, so cavity effects only enter through the bilinear coupling and dipole self-energy components of the Hamiltonian. As such, we expect cavity-induced changes to the electronic structure to be insensitive to the choice of electronic XC functional. Table \ref{TAB:SVWN3_PBE} provides computed \textit{de} values from relaxed QED-TDDFT calculations performed using the SVWN3 and PBE functionals and the 6-31G(d,p) basis. We first note that the cavity-free predictions for the \textit{de} are in qualitative agreement with those from B3LYP and experiment;\cite{Hanson18_28478} the (\textit{S})-Boc-Pro-(\textit{R})-BINOL diastereomer is favored. However, while PBE-derived results (15\% \textit{de}) are in good quantitative agreement with those from B3LYP (17\% \textit{de}), SVWN3 predicts a slightly larger preference for (\textit{S})-Boc-Pro-(\textit{R})-BINOL (39\% \textit{de}). Second, we note that the same general trends with respect to cavity mode polarization axis and coupling strength observed in Table \ref{TAB:B3LYP_RELAXED} are present in the SVWN3 and PBE data. For a $z$-polarized cavity mode, all three functionals predict an increase in the preference for the (\textit{S})-Boc-Pro-(\textit{R})-BINOL diastereomer, and this increase amounts to as much as 25 percentage points in the case of PBE (15\% to 40\% \textit{de}) and 19 percentage points for SVWN (29\% to 58\% \textit{de}). When the cavity mode is polarized along the molecular dipole moment (as predicted by SVWN3 or PBE), the preferred diastereomer can change. PBE predicts that at $\lambda = 0.05$, there will be a 40\% \textit{de} of (\textit{S})-Boc-Pro-(\textit{S})-BINOL [a 55 percentage point swing from 15\%(\textit{R})], whereas SVWN3 predicts that there will be 5\% \textit{de} of (\textit{S})-Boc-Pro-(\textit{S})-BINOL at this coupling strength [a 44 percentage point swing from 39\%(\textit{R})]. The magnitudes of the swing from a preference for (\textit{S})-Boc-Pro-(\textit{R})-BINOL to one for (\textit{S})-Boc-Pro-(\textit{S})-BINOL are comparable to that which was observed for B3LYP in Table \ref{TAB:B3LYP_RELAXED}. Also, as was observed for the case of B3LYP, the \textit{de} is not particularly sensitive to the presence of the cavity when the cavity mode is polarized in the $x$- or $y-$ directions. 

	\begin{table}
		\begin{threeparttable}
			\caption {Computed diastereomeric excess for (\textit{S})-Boc-Pro-(\textit{R}/\textit{S})-BINOL from PBE and SVWN3 functionals with $\omega_{\rm cav} = 10.0$ eV. The \textit{de} was determined according to the relative energies of the first electronic excited states of deprotonated (\textit{S})-Boc-Pro-(\textit{R}/\textit{S})-BINOL. } \label{TAB:SVWN3_PBE}
			\footnotesize\centering
			\setlength{\tabcolsep}{0em}
			\begin{tabular*}{\linewidth}{@{\extracolsep{\fill}}
					lccccc}
			
			& & \multicolumn{3}{c}{mode polarization} \\
			\cline{3-5}
			&$\lambda$& $x$ & $y$ & $z$ &dipole \\
\midrule
                & 0.00 & 15(\textit{R}) & 15(\textit{R}) & 15(\textit{R}) & 15(\textit{R})   \\
                & 0.01 & 14(\textit{R}) & 15(\textit{R}) & 16(\textit{R}) & 12(\textit{R})   \\
PBE             & 0.02 & 13(\textit{R}) & 14(\textit{R}) & 19(\textit{R}) &  3(\textit{R})   \\
                & 0.03 & 11(\textit{R}) & 13(\textit{R}) & 24(\textit{R}) & 11(\textit{S})   \\
                & 0.04 &  9(\textit{R}) & 12(\textit{R}) & 32(\textit{R}) & 25(\textit{S})   \\
                & 0.05 &  6(\textit{R}) & 11(\textit{R}) & 40(\textit{R}) & 40(\textit{S})   \\

\midrule
                & 0.00 & 39(\textit{R}) & 39(\textit{R}) & 39(\textit{R}) & 39(\textit{R})   \\
                & 0.01 & 38(\textit{R}) & 38(\textit{R}) & 39(\textit{R}) & 36(\textit{R})   \\
SVWN3           & 0.02 & 37(\textit{R}) & 38(\textit{R}) & 42(\textit{R}) & 29(\textit{R})   \\
                & 0.03 & 35(\textit{R}) & 37(\textit{R}) & 47(\textit{R}) & 19(\textit{R})   \\
                & 0.04 & 33(\textit{R}) & 36(\textit{R}) & 52(\textit{R}) &  7(\textit{R})   \\
                & 0.05 & 31(\textit{R}) & 35(\textit{R}) & 58(\textit{R}) &  5(\textit{S})   \\
		\end{tabular*}
	\end{threeparttable}
\end{table}

\begin{figure}[!htpb]
	\includegraphics{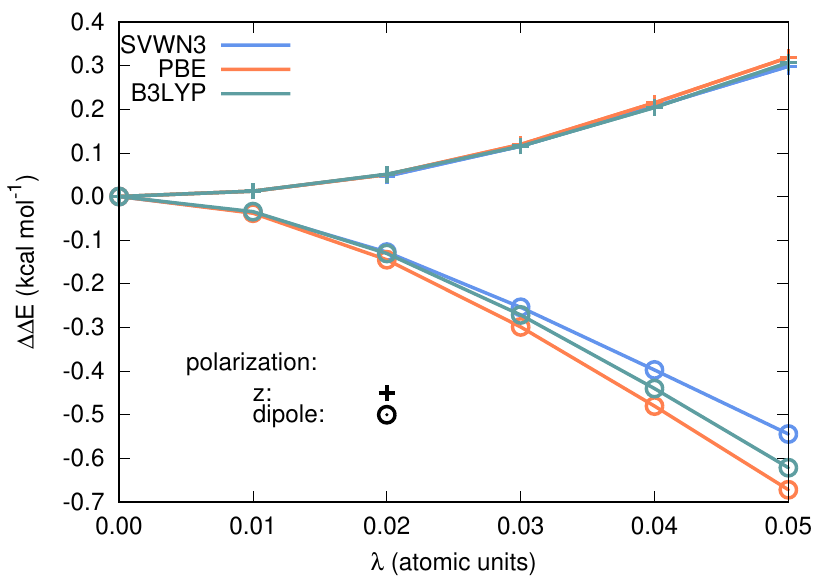}
	\caption{Change in the energy difference between the (\textit{S})-Boc-Pro-(\textit{R}/\textit{S})-BINOL diastereomers ($\Delta \Delta E$, kcal mol$^{-1}$) as a function of coupling strength for cavity modes with $\omega_{\rm cav} = 10.0$ eV polarized along the $z$-direction and the direction defined by the molecular dipole moment for each diastereomer.}\label{FIG:DELTA_DELTA_E}
\end{figure}

The similarities between results obtained from each functional are more easily visualized in Fig. \ref{FIG:DELTA_DELTA_E}, which depicts the change in the predicted energy gap between the (\textit{S})-Boc-Pro-(\textit{R}/\textit{S})-BINOL diastereomers for each functional ($\Delta \Delta E$), when considering cavity modes with $\omega_{\rm cav} = 10.0$ eV. We focus on the cases for which the most dramatic changes in the \textit{de} are observed: cavity modes polarized in the $z$-direction or along the molecular dipole moments. At all coupling strengths, the B3LYP, PBE, and SVWN3 functionals provide comparable results. For the largest coupling strength considered ($\lambda = 0.05$), the relative energies of the (\textit{S})-Boc-Pro-(\textit{R}/\textit{S})-BINOL diastereomers change by $\approx$ 0.30 to 0.32 kcal mol$^{-1}$ or $\approx$ -0.54 to -0.67 kcal mol$^{-1}$ when the cavity mode is polarized in the $z$-direction or along the molecular dipole moments, respectively. We can attribute the larger spread in $\Delta \Delta E$ when the cavity is polarized along the dipole moments of the diastereomers to the fact that the precise orientations of the dipole moments differ slightly at each level of theory. Nonetheless, the qualitative behavior of each functional is similar.



\section{Conclusions}
\label{SEC:CONCLUSIONS}

We have demonstrated computationally that strong light-matter interactions can be leveraged to discriminate between diastereomers, within the specific context of ESPT-driven enantioenrichment of BINOL. Changes to both the attainable \textit{de} and the handedness of BINOL are achievable, but the outcomes are highly sensitive to the relative orientation of the molecules and the cavity mode polarization axis (see the molecular orientation in Fig.~\ref{FIG:MOLECULAR_ORIENTATION}). For example, a $z$-polarized cavity mode serves to enhance the predicted \textit{de} for the cavity-free favored diastereomer [(\textit{S})-Boc-Pro-(\textit{R})-BINOL] from 17\% to $\approx$ 40\%, while a cavity mode polarized along the molecular dipole moment results in a reversed preference for (\textit{S})-Boc-Pro-(\textit{S})-BINOL with predicted \textit{de} as large as 34\%. These outcomes are intriguing as they suggest that precise control over molecular orientation within an optical cavity can enable the targeted generation of a preferred diastereomer without the need to change the chiral directing group. Moreover, we note that the observed diastereomer discrimination is achieved via strong coupling to \textit{linearly}-polarized cavity modes. In this way, our work complements other recent work on {\em chiral} optical cavities\cite{Koch22_chiral} in which circularly-polarized cavity modes have been shown to be effective at discriminating between enantiomers of chiral molecules. 

We have found that the observed cavity effect is insensitive to the mode frequency, which suggests that it can be attributed to cavity-induced modifications to the ground-state electronic structure. Indeed, we only observe meaningful changes to the \textit{de} when the ground-state Kohn-Sham orbitals are allowed to relax within the cavity. Such ground-state effects derive entirely from the dipole-self energy component of the Hamiltonian and only become important in strong single-molecule coupling scenarios. Hence, experimental realization of the effects that we have predicted will require cavity architectures that support either a single photon mode with a few- or sub-nm$^{3}$ volume,\cite{Baumberg16_127} or multiple modes polarized along the same direction. For ground-states described by mean-field cavity QED methods (such as QED-DFT), the latter case corresponds to a cavity QED simulation in which the single-mode coupling strength is replaced by an effective coupling deriving from the cumulative effect of multiple modes ({\em i.e.}, $\lambda_{\rm eff}^2 = \sum_i \lambda^2_i$).\cite{Koch21_094113}




\section{Appendix A: Ground-state QED-DFT}

In this appendix, we provide additional details regarding our formulation of the QED-Kohn-Sham ground-state problem. Given the Pauli-Fierz Hamiltonian in Eq.~\ref{EQN:HPF_COHERENT}, which is represented within the coherent-state basis, the ground-state QED-DFT energy is
\begin{eqnarray}
\label{EQN:QED_DFT_ENERGY}
    E &=& \sum_{\mu\nu} ( T_{\mu\nu} + V_{\mu\nu} + \frac{1}{2} J_{\mu\nu} ) \gamma_{\rm \mu\nu}  
    + E_{\rm xc}[\rho_\alpha, \rho_\beta, ...] \nonumber \\
    &+& \langle \frac{1}{2} [{\bm{\lambda}} \cdot ({\bm{\hat{\mu}}} - \langle {\bm{\hat{\mu}}} \rangle)]^2 \rangle
\end{eqnarray}
where, $\mu$ and $\nu$ index atomic basis functions, 
$T_{\mu\nu}$ and $V_{\mu\nu}$ are electron kinetic energy and electron-nucleus potential energy integrals, respectively, and $J_{\mu\nu}$ represents the coulomb matrix
\begin{equation}
    J_{\mu\nu} = \sum_{\lambda \sigma} \langle\mu\lambda|\nu\sigma\rangle \gamma_{\lambda\sigma}
\end{equation}
Here, $\langle\mu\lambda|\nu\sigma\rangle$ is a two-electron repulsion integral in physicists' notation. 
The symbol $\gamma_{\mu\nu}$ represents an element of the one-particle reduced-density matrix
\begin{equation}
    \gamma_{\mu\nu} = \sum_i^{N_{\rm e}} c^*_{\mu i} c_{\nu i}
\end{equation}
where $N_{\rm e}$ is the number of electrons and $\{c_{\mu i}\}$ are  molecular orbital coefficients.
$E_{\rm xc}$ is a standard electron exchange-correlation functional that depends upon the $\alpha$- and $\beta$-spin densities ($\rho_\alpha$ and $\rho_\beta$, respectively), as well as additional quantities, depending on the density functional approximation.  The last term in Eq.~\ref{EQN:QED_DFT_ENERGY} is the dipole self-energy. Note that, for mean-field methods, the bilinear coupling term in Eq.~\ref{EQN:HPF_COHERENT} does not contribute to the total energy when the Hamiltonian is represented in the coherent-state basis. Note also that this energy expression could in principle be modified to  include some fraction of exact Hartree-Fock exchange.

To arrive at a useful expression for the dipole self-energy, we first note that, in the coherent-state basis and within the cavity Born-Oppenheimer approximation, the operator $\bm{\hat{\mu}}-\braket{\bm{\hat{\mu}}}$ consists of only electronic components, {\em i.e.},
\begin{equation}
    {\bm {\hat{\mu}}} - \langle {\bm {\hat{\mu}}} \rangle = {\bm{\hat{\mu}}}_{\rm e} - \langle {\bm{\hat{\mu}}}_{\rm e} \rangle
\end{equation}
Now, the dipole-self energy operator takes the form
\begin{eqnarray}
\label{EQN:DSE_OPERATOR}
    \frac{1}{2} [{\bm{\lambda}} \cdot ({\bm{\hat{\mu}}}_e - \langle {\bm{\hat{\mu}}}_e \rangle)]^2 &=& \frac{1}{2} ( {\bm{\lambda}}\cdot  {\bm{\hat{\mu}}}_{\rm e} ) ^2 \nonumber \\
    &-& \nonumber ( {\bm{\lambda}}\cdot {\bm{\hat{\mu}}}_{\rm e} ) ( {\bm{\lambda}}\cdot \langle {\bm{\hat{\mu}}}_{\rm e}\rangle ) \\
    &+& \frac{1}{2} ( {\bm{\lambda}}\cdot \langle {\bm{\hat{\mu}}}_{\rm e}\rangle ) ^2
\end{eqnarray}
The dipole-squared operator can be expanded in terms of one- and two-electronic contributions as
\begin{equation}
    ( {\bm{\lambda}}\cdot  {\bm{\hat{\mu}}}_{\rm e} ) ^2 = \sum_{i \neq j} [ {\bm{\lambda}}\cdot  {\bm{\hat{\mu}}}_{\rm e}(i) ][ {\bm{\lambda}}\cdot  {\bm{\hat{\mu}}}_{\rm e}(j)] + \sum_i [ {\bm{\lambda}}\cdot  {\bm{\hat{\mu}}}_{\rm e}(i) ]^2
\end{equation}
In second-quantized notation, this operator has the form
\begin{eqnarray}
\label{EQN:DIPOLE_SQUARED}
    ( {\bm{\lambda}}\cdot  {\bm{\hat{\mu}}}_{\rm e} ) ^2 &=&  \sum_{\mu\nu\lambda\sigma} d_{\mu\nu} d_{\lambda\sigma} \hat{a}^\dagger_\mu \hat{a}^\dagger_\lambda \hat{a}_\sigma \hat{a}_\nu \nonumber \\
    &-& \sum_{\mu\nu} q_{\mu\nu} \hat{a}^\dagger_\mu \hat{a}_\nu
\end{eqnarray}
The symbols $\hat{a}^\dagger$ and $\hat{a}$ represent fermionic creation and annihilation operators, respectively, $d_{\mu\nu}$ represents a modified dipole integral
\begin{equation}
    d_{\mu\nu} = - \sum_{a \in \{x,y,z\}} \lambda_a \int \chi^*_\mu r_a \chi_{\nu} d\tau
\end{equation}
and $q_{\mu\nu}$ is a modified quadrupole integral
\begin{equation}
    q_{\mu\nu} = - \sum_{ab \in \{x,y,z\}} \lambda_a \lambda_b \int \chi^*_\mu r_a r_b \chi_{\nu} d\tau
\end{equation}
Here, $\chi_\mu$ represents an atomic basis function, $\lambda_a$ is a cartesian component of ${\bm{\lambda}}$, and $r_a$ is a cartesian component of the position vector ({\em e.g.}, for ${\mathbf{r}} = (x, y, z)$, $r_x$ = $x$).
Given Eq.~\ref{EQN:DIPOLE_SQUARED} and 
\begin{equation}
    ( {\bm{\lambda}}\cdot {\bm{\hat{\mu}}}_{\rm e} ) = \sum_{\mu\nu} d_{\mu\nu} \hat{a}^\dagger_\mu \hat{a}_\nu
\end{equation}
we arrive at the final second-quantized form of Eq.~\ref{EQN:DSE_OPERATOR}
\begin{eqnarray}
\label{EQN:DSE_FINAL}
\frac{1}{2} [{\bm{\lambda}} \cdot ({\bm{\hat{\mu}}}_e - \langle {\bm{\hat{\mu}}}_e \rangle)]^2 &=& \frac{1}{2} \sum_{\mu\nu\lambda\sigma} d_{\mu\nu} d_{\lambda\sigma} \hat{a}^\dagger_\mu \hat{a}^\dagger_\lambda \hat{a}_\sigma \hat{a}_\nu \nonumber \\ 
&+& \sum_{\mu\nu} O^{\rm DSE}_{\mu\nu} \hat{a}^\dagger_\mu \hat{a}_\nu + \frac{1}{2} ({\bm{\lambda}}\cdot\langle {\bm{\mu}}_{\rm e}\rangle)^2
\end{eqnarray}
where 
\begin{equation}
    O^{\rm DSE}_{\mu\nu} = -( {\bm{\lambda}}\cdot \langle {\bm{\hat{\mu}}}_{\rm e}\rangle ) d_{\mu\nu} - \frac{1}{2} q_{\mu\nu}  
\end{equation}
For a single Slater determinant, the expectation value of Eq.~\ref{EQN:DSE_FINAL} is
\begin{widetext}
\begin{equation}
    \langle \frac{1}{2} [{\bm{\lambda}} \cdot ({\bm{\hat{\mu}}}_e - \langle {\bm{\hat{\mu}}}_e \rangle)]^2 \rangle = \frac{1}{2} \sum_{\mu\nu\lambda\sigma} d_{\mu\nu} d_{\lambda\sigma} ( \gamma_{\mu\nu} \gamma_{\lambda\sigma} - \gamma_{\mu\sigma}\gamma_{\lambda\nu}) + \sum_{\mu\nu}O^{\rm DSE}_{\mu\nu} \gamma_{\mu\nu}+ \frac{1}{2} ({\bm{\lambda}}\cdot\langle {\bm{\mu}}_{\rm e}\rangle)^2
\end{equation}
\end{widetext}
or
\begin{eqnarray}
    \langle \frac{1}{2} [{\bm{\lambda}} \cdot ({\bm{\hat{\mu}}}_e - \langle {\bm{\hat{\mu}}}_e \rangle)]^2 \rangle &=& \sum_{\mu\nu} (\frac{1}{2} J^{\rm DSE}_{\mu\nu} - \frac{1}{2} K^{\rm DSE}_{\mu\nu} + O^{\rm DSE}_{\mu\nu} ) \gamma_{\mu\nu} \nonumber \\
    &+& \frac{1}{2} ({\bm{\lambda}}\cdot\langle {\bm{\mu}}_{\rm e}\rangle)^2
\end{eqnarray}
where
\begin{equation}
    J^{\rm DSE}_{\mu\nu} = d_{\mu\nu} \sum_{\lambda \sigma} d_{\lambda \sigma} \gamma_{\lambda\sigma}
\end{equation}
and
\begin{equation}
    K^{\rm DSE}_{\mu\nu} = \sum_{\lambda \sigma} d_{\mu\sigma} d_{\lambda \nu} \gamma_{\lambda\sigma}
\end{equation}

With all of the components of the energy (Eq.~\ref{EQN:QED_DFT_ENERGY}) defined, we can make this energy stationary with respect to the orbitals that define the spin densities and density matrix, while enforcing orthogonality of these orbitals. Doing so results in a set of Kohn-Sham equations that resembles those in the ordinary electronic problem, except for the presence of dipole self-energy contributions. In other words, the orbitals in Kohn-Sham QED-DFT are determined as eigenfunctions of the modified Kohn-Sham matrix
\begin{eqnarray}
    F^{\rm KS}_{\mu\nu} &=& T_{\mu\nu} + V_{\mu\nu} + J_{\mu\nu} + V^{\rm xc}_{\mu\nu} \nonumber \\
    &+& O^{\rm DSE}_{\rm \mu\nu} + J_{\mu\nu}^{\rm DSE} - K_{\mu\nu}^{\rm DSE} 
\end{eqnarray}
where $V^{\rm xc}_{\mu\nu}$ represents the electron exchange-correlation potential matrix. Of course, this Kohn-Sham matrix could be modified to include some fraction of exact Hartree-Fock exchange.

\section{Appendix B: QED-RPA equations}
In this appendix, we derive a cavity QED formulation of the random phase approximation (RPA). QED-TDDFT equations are closely related to the QED-RPA ones, as is the case with standard TDDFT and RPA. We approach this derivation from the point of view of Rowe's equation of motion.\cite{Rowe68_153}

Consider an excited-state wave function
\begin{equation}
    |\Psi_n\rangle = \hat{O}^\dagger_n |0_{\rm e}0_{\rm p}\rangle
\end{equation}
and the associated Schr\"{o}dinger equation
\begin{equation}
    \hat{H} \hat{O}^\dagger_n |0_{\rm e}0_{\rm p}\rangle = E_n \hat{O}^\dagger_n |0_{\rm e}0_{\rm p}\rangle
\end{equation}
Assuming  $|0_{\rm e}0_{\rm p}\rangle = |0_{\rm e}\rangle \otimes |0_{\rm p}\rangle$ is an eigenfunction of $\hat{H}$, we have
\begin{equation}
\label{EQN:SE_SINGLE_COMMUTATOR}
    \langle 0_{\rm e}0_{\rm p} | \hat{A}  [\hat{H}, \hat{O}^\dagger_n] |0_{\rm e}0_{\rm p}\rangle = \Omega_n  \langle 0_{\rm e}0_{\rm p} | \hat{A}   \hat{O}^\dagger_n |0_{\rm e}0_{\rm p}\rangle
\end{equation}
where $\Omega_n = E_n - E_0$, $E_0$ is the energy associated with $|0_{\rm e}0_{\rm p}\rangle$, and $\hat{A}$ is defined such that $\hat{A}^\dagger|0_{\rm e}0_{\rm p}\rangle$ is an arbitrary state within the manifold of states spanned by $\hat{O}^\dagger_n |0_{\rm e}0_{\rm p}\rangle$. If $\hat{O}^\dagger_n$ satisfies the killer condition, $\hat{O}_n | 0_{\rm e}0_{\rm p}\rangle = 0$, then Eq.~\ref{EQN:SE_SINGLE_COMMUTATOR} is equivalent to 
\begin{equation}
\label{EQN:SE_DOUBLE_COMMUTATOR}
    \langle 0_{\rm e}0_{\rm p} | [\hat{A},  [\hat{H}, \hat{O}^\dagger_n]] |0_{\rm e}0_{\rm p}\rangle = \Omega_n  \langle 0_{\rm e}0_{\rm p} | [\hat{A},   \hat{O}^\dagger_n] |0_{\rm e}0_{\rm p}\rangle
\end{equation}

To obtain the working equations for QED-RPA, we define an approximate transition operator (which does not actually satisfy the killer condition) as
\begin{equation}
\label{EQN:RPA_EXPANSION}
    \hat{O}_n^\dagger = \sum_{ia} (X_{ai}^n \hat{X}_{ai}^\dagger + Y_{ai}^n \hat{Y}_{ai}^\dagger ) + M^n \hat{M}^\dagger  + N^n \hat{N}^\dagger
\end{equation}
with
\begin{eqnarray}
\hat{X}_{ai}^\dagger &=& \hat{a}^\dagger_a \hat{a}_i \\
\hat{Y}_{ai}^\dagger &=& -\hat{a}^\dagger_i \hat{a}_a \\
\hat{M}^\dagger &=& \hat{b}^\dagger \\
\hat{N}^\dagger &=& -\hat{b}
\end{eqnarray}
Here, the labels $i$ and $a$ represent occupied and virtual molecular orbitals, respectively.
The expansion coefficients in Eq.~\ref{EQN:RPA_EXPANSION} ($X^n_{ai}$, $Y^n_{ai}$, $M^n$, and $N^n$) can be determined as the eigenvectors of the
generalized eigenvalue problem given in Eq.~\ref{EQN:QED_TDDFT_EQUATIONS} with the blocks on the left-hand side of that equation defined by
\begin{align}
       \langle 0_{\rm e}0_{\rm p} |~[\hat{X}_{ai}, ~[\hat{H}, ~\hat{X}_{bj}^\dagger]]~| 0_{\rm e}0_{\rm p} \rangle &= ({\bm A} + {\bm \Delta})_{ai,bj} \\
        \langle 0_{\rm e}0_{\rm p} |~[\hat{X}_{ai}, ~[\hat{H}, ~\hat{Y}_{bj}^\dagger]]~| 0_{\rm e}0_{\rm p} \rangle &= ({\bm B} + {\bm \Delta^\prime})_{ai,bj} \\
\langle 0_{\rm e}0_{\rm p} |~[\hat{X}_{ai}, ~[\hat{H}, ~\hat{M}^\dagger]]~| 0_{\rm e}0_{\rm p} \rangle &= g_{ai} \\
        \langle 0_{\rm e}0_{\rm p} |~[\hat{X}_{ai}, ~[\hat{H}, ~\hat{N}^\dagger]]~| 0_{\rm e}0_{\rm p} \rangle &= g_{ai} \\
        \langle 0_{\rm e}0_{\rm p} |~[\hat{Y}_{ai}, ~[\hat{H}, ~\hat{Y}_{bj}^\dagger]]~| 0_{\rm e}0_{\rm p} \rangle &= ({\bm A} + {\bm \Delta})_{ai,bj} \\
        \langle 0_{\rm e}0_{\rm p} |~[\hat{Y}_{ai}, ~[\hat{H}, ~\hat{M}^\dagger]]~| 0_{\rm e}0_{\rm p} \rangle &= g_{ai} \\
        \langle 0_{\rm e}0_{\rm p} |~[\hat{Y}_{ai}, ~[\hat{H}, ~\hat{N}^\dagger]]~| 0_{\rm e}0_{\rm p} \rangle &= g_{ai} \\
        \langle 0_{\rm e}0_{\rm p} |~[\hat{M}, ~[\hat{H}, ~\hat{M}^\dagger]]~| 0_{\rm e}0_{\rm p} \rangle &= \omega_{\rm cav} \\
        \langle 0_{\rm e}0_{\rm p} |~[\hat{M}, ~[\hat{H}, ~\hat{N}^\dagger]]~| 0_{\rm e}0_{\rm p} \rangle &= 0 \\
        \langle 0_{\rm e}0_{\rm p} |~[\hat{N}, ~[\hat{H}, ~\hat{N}^\dagger]] ~| 0_{\rm e}0_{\rm p} \rangle &= \omega_{\rm cav}
\end{align}
etc. The blocks on the right-hand side of Eq.~\ref{EQN:QED_TDDFT_EQUATIONS} can similarly be defined using the appropriate single commutators (see the right-hand side of Eq.~\ref{EQN:SE_DOUBLE_COMMUTATOR}).
Above,  
\begin{equation}
g_{ai} = \sqrt{\frac{\omega_{\rm cav}}{2}} d_{ai} 
\end{equation}
and ${\bm A}$ and ${\bm B}$ are the standard matrices that arise in RPA, {\em i.e.}, 
\begin{eqnarray}
A_{ai,bj} &=& \delta_{ab}\delta_{ij}(\epsilon_a - \epsilon_i) + \langle aj || ib \rangle \\
B_{ai,bj} &=& \langle ab || ij \rangle
\end{eqnarray}
Here, $\epsilon_i$ and $\epsilon_a$ are orbital energies, and $\langle aj || ib \rangle = \langle aj | ib \rangle - \langle aj | bi \rangle$ is an antisymmetrized two-electron integral in physicists' notation.
Dipole self-energy contributions arise in two places here. First, one-electron components are contained within the orbital energies (if the underlying QED-DFT procedure relaxes the Kohn-Sham orbitals to account for the presence of the cavity). Second,
the tensors ${\bm \Delta}$ and ${\bm \Delta^\prime}$ contain two-electron contributions, and the form of these contributions is 
similar to that in the ${\bm A}$ and ${\bm B}$ matrices:
\begin{eqnarray}
\label{EQN:DELTA}
\Delta_{ai,bj} = d_{ai} d_{jb} - d_{ab} d_{ij} \\
\label{EQN:DELTA_PRIME}
\Delta^\prime_{ai,bj} = d_{ai} d_{bj} - d_{aj} d_{ib}
\end{eqnarray}

The QED-TDDFT equations we solve in this work are obtained by replacing the exchange parts of the ${\bm A}$ and ${\bm B}$ matrices with appropriate derivatives of the exchange-correlation functional. Given this form, there are a few important distinctions between our approach and that outlined in Ref.~\citenum{Shao21_064107}. First, our reference state, $|0_{\rm e}0_{\rm p}\rangle$, is determined via the modified Kohn-Sham procedure described in Sec.~\ref{SEC:THEORY}, which includes cavity interactions, whereas the ground-state in Ref.~\citenum{Shao21_064107} is an unperturbed Kohn-Sham state. Cavity effects in the ground-state calculation will affect the orbitals and orbital energies entering the QED-TDDFT procedure. Second, the Hamiltonian entering our QED-TDDFT procedure is represented in the coherent-state basis (Eq.~\ref{EQN:HPF_COHERENT}), whereas that used in Ref.~\citenum{Shao21_064107} is not. Third, our ${\bm \Delta}$ and ${\bm \Delta^\prime}$ matrices contain exchange-like contributions (the second terms in Eqs.~\ref{EQN:DELTA_PRIME} and \ref{EQN:DELTA_PRIME}), whereas Ref.~\citenum{Shao21_064107} ignores these quantities. When ignoring these exchange-like contributions, ${\bm \Delta} = {\bm \Delta^\prime}$, if the molecular orbitals are real-valued.

\vspace{0.5cm}

\noindent {\bf Supporting Information:} Energies (E$_{\rm h}$) of the first electronic excited states of deprotonated (\textit{S})-Boc-Pro-(\textit{R})-BINOL and (\textit{S})-Boc-Pro-(\textit{S})-BINOL molecules coupled to a 10 eV cavity mode computed at the relaxed QED-TDDFT/6-31G(d,p) level of theory using the PBE, B3LYP, and SVWN3 functionals.

\vspace{0.5cm}

\begin{acknowledgments}
This material is based upon work supported by the National Science Foundation under Grant No.~CHE-2100984.
\end{acknowledgments}

\bibliography{Journal_Short_Name.bib,cqed.bib,other.bib,binol.bib}

\end{document}